\documentclass[a4paper,11pt]{article}
\pdfoutput=1 

\usepackage{jinstpub} 

\usepackage{graphicx,color,footnote}
\newcommand{\wc}{\emph{Wire-Cell}}

\title{Three-dimensional Imaging for Large LArTPCs}

\author[1]{X. Qian,\note{Corresponding author}}
\author[2]{C. Zhang, \note{Corresponding author}}
\author{B. Viren,}
\author{and M. Diwan}
\affiliation{Physics Department, Brookhaven National Laboratory, \\
Upton, NY, 11973 U.S.A.}
\emailAdd{xqian@bnl.gov}
\emailAdd{chao@bnl.gov}

\abstract{
  High-performance event reconstruction is critical for current and
  future massive liquid argon time projection chambers (LArTPCs) to realize their full scientific potential. LArTPCs with readout using wire planes provide a limited number of 2D projections. 
  In general,  without a pixel-type readout it is challenging to achieve unambiguous 3D event reconstruction. 
  As a remedy, we present a novel 3D imaging method, \wc,
  which incorporates the charge and sparsity information in addition
  to the time and geometry through simple and robust
  mathematics. The resulting 3D image of ionization density provides an excellent starting point for further reconstruction and enables the true power of 3D tracking
  calorimetry in LArTPCs.
}


\begin{document}

\maketitle
\flushbottom

\section{Introduction} 
Development of large fine-grained neutrino detectors with excellent particle identification and energy resolution is motivated by the science of neutrino oscillations~\cite{pdg-nakamura,Diwan:2016gmz}. 
The Liquid Argon Time Projection Chamber (LArTPC) has been shown to be an attractive detector technology because of its low cost, high density, long lifetime for ionization electrons, and low electron diffusion ~\cite{lartpc,lartpc_3,Willis:1974gi,Nygren:1976fe}.  
Experiments on many scales (from a few liters to hundreds of tons) have been built to optimize the design of LArTPCs for neutrino detection~\cite{Anderson:2011ce,Cavanna:2014iqa,Berns:2013usa,Acciarri:2016smi,Amerio:2004ze}.  
Demonstration of  both reliable hardware performance and efficient reconstruction of neutrino events with well-identified leptonic final states is very important for the future Fermilab short-baseline neutrino program (SBN)~\cite{Antonello:2015lea} and long-baseline deep underground neutrino experiments (DUNE)~\cite{Acciarri:2016crz}.
As a high-resolution tracking calorimeter, LArTPC should allow for detailed reconstruction of the trajectories and energy deposition of charged particles. These aspects are critical for DUNE to fulfill its potential in searching for leptonic CP violation~\cite{Diwan:2016gmz}, determining neutrino mass hierarchy~\cite{Qian:2015waa}, performing precision tests of the three-neutrino model~\cite{Qian:2013ora}, and other physics~\cite{Adams:2013qkq}. In this paper we analyze methods for achieving high-performance event reconstruction which currently remains an open challenge.

The single-phase wire-plane based design for a LArTPC has cathode and anode planes separated by a long drift-distance to create an electric field in the large volume of liquid argon.  The ionization electrons from charged particles drift along the electric field toward the anode planes.  Their motion induces a current in the wires which are read out by low noise electronics~\cite{Radeka:2011zz}. 
Multiple anode planes, each with parallel wires oriented at different angles, allow three-dimensional reconstruction of the ionization energy depositions. This design is a relatively cost-effective way to build a cryogenic detector with the required large mass.  The spacing between wires can be small (a few mm) to get excellent sampling of tracks and electromagnetic showers.
In contrast to a pixel-based readout used by collider gas TPCs, the wire-plane readout gives rise to reconstruction difficulties and ambiguities because of the projective geometry.
There is naturally information loss in going from $O(n^2)$ pixels to $O(n)$ wires.  
For tracks that have
small angles to the plane perpendicular to the electric field, the ionization electrons from the entire track produce simultaneous pulses in all wires in their drift path.
Such \emph{isochronous} events are difficult to reconstruct, as the ambiguities grow exponentially with the multiplicity of simultaneously hit wires.
In addition, isochronous conditions are often met for large electromagnetic
showers, and near particle interaction vertices where track multiplicity is high.

In the usual approach, reconstruction is performed on each 2D wire-versus-time view separately (see e.g. Ref.~\cite{Acciarri:2017hat}).
After pattern recognition, the clustered 2D objects are matched from the different views to form 3D objects in order to determine track angles and $dE/dx$. This approach becomes challenging when objects that are well separated in 3D yet overlap in one or more of the 2D views. 
In addition, the intrinsic ambiguities for isochronous events often cannot be reduced without applying early heuristics which make assumptions about the final event topology. 

Here we describe a new 3D imaging method, \wc, which takes advantage of key features of the LArTPC to directly reconstruct the ionization density in 3D voxels. In order to resolve many of the ambiguities caused by the lack of pixel-level information, the \wc{} method incorporates the measured ionization charge information and its sparsity in addition to its arrival time and the detector wire geometry through simple and robust mathematics.  Similar to the pixel-readout
detectors, the resulting clean and accurate 3D image provides an excellent starting point for subsequent pattern recognition procedures.

\section{Wire-Cell 3D Imaging}

\subsection{Description of Basic Principles}
The key concept of \wc{} is to tomographically reconstruct the 3D image of ionization density per time-slice regardless of event topology. This approach utilizes the unique feature of the LArTPC that the same number of drifting electrons (``charge'') is measured redundantly by all wire planes. More explicitly, the early wire planes receive bipolar induction signals as ionization electrons pass by, while the same electrons are collected on the final wire plane leading to unipolar induction signals. The imaging procedure has the following steps:

\paragraph*{Signal preparation: }
The bipolar and unipolar signals that come from the low-noise shaping amplifiers are first filtered with standard techniques to reduce excess noise~\cite{Acciarri:2017sde} and then processed to statistically remove the induction response shape~\cite{Baller:2017ugz,Adams:2018dra}. The end result is a normalized pulse of charge for each wire. After this non-trivial signal processing procedure, excellent signal matching among different wire planes  has been demonstrated~\cite{Adams:2018gbi}. This establishes the basis of \wc.
Event displays exhibiting the effectiveness of the overall signal processing chain
can be found in Ref.~\cite{magnify_online}.

\begin{figure}
\centering
\includegraphics[width=0.8\textwidth,clip,trim=10mm 12mm 9mm 9mm]{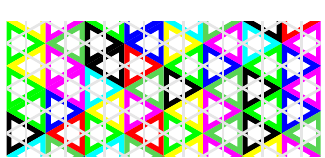}
\caption{\label{fig:cells} Cells constructed with the MicroBooNE detector geometry.
  Cell boundaries are represented by colored lines, while the wire centers are represented by
  gray lines. All cells have equilateral triangular shapes due to the $\pm60^{\circ}$ wire orientations.}
\end{figure}

The next step is to divide the signals into time-slices. Each time slice represents a tomographic cross
section across the detector with its normal in the direction of the nominal E-field. The width of the time slice,
typically $\sim$2 $\mu$s, is chosen to be consistent with
the digitization sampling rate, electronics shaping, and the expected diffusion~\cite{Li:2015rqa}. 
For each time slice, the algorithm identifies all possible ``hit cells'' that correspond to all
the wires with sizable signals (namely ``hit wires''). A \emph{wire} is defined as a 2D
region centered around the wire location with the width equal to the wire pitch.
A \emph{cell} is then defined as the overlapping region formed by the
nearest wire from each plane. Figure~\ref{fig:cells} shows the pattern of
cells from the MicroBooNE detector geometry~\cite{Acciarri:2016smi}. Based on the hit
wires, the possible hit cells are constructed.  Any ambiguities at this point are simply retained and addressed by
the following steps.

\begin{figure}
	\centering
        \includegraphics[width=0.8\textwidth]{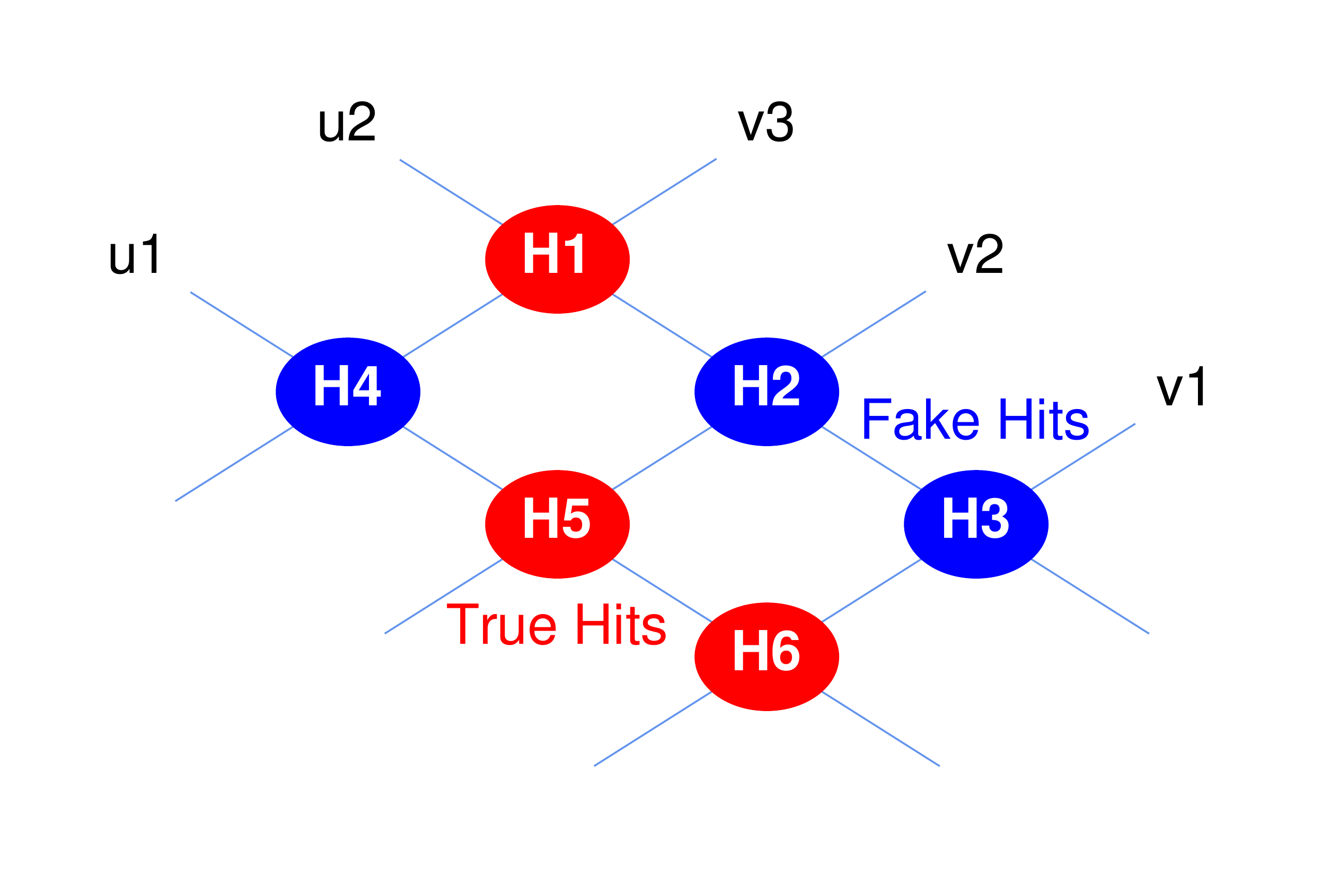}
	\caption{\label{fig:ambiguity} Illustration of the hit ambiguity caused by
          the wire readout in a simplified two-wire-plane example. }
\end{figure}

\paragraph*{Incorporating charge information: }
Voxel-based reconstruction is natural in pixel-readout detectors
but it is hindered in LArTPC because of the ambiguities due to the limited number of angular views provided by the wire planes. Such
ambiguities are illustrated in Fig.~\ref{fig:ambiguity} with a simplified example consisting of only two wire planes.  There, three cells (\textsf{H1}, \textsf{H5}, \textsf{H6}) are ``hit'' by the distribution of charge
passing through the wire planes in the given single time slice.
Due to the wire readout, five wires (\textsf{u1}, \textsf{u2}, \textsf{v1}, \textsf{v2}, \textsf{v3})
record signals. This leads to 6 possible hit cells, including 3 fake ones (\textsf{H2}, \textsf{H3}, \textsf{H4}).
Such ambiguities cannot be resolved with geometry information alone.

However, given that any charge deposited inside a cell is measured independently by
the associated wires, additional charge equations can be constructed. For the example in Fig.~\ref{fig:ambiguity}, we have:
\begin{equation}
\left( \begin{array}{c} \mathsf{u1} \\ \mathsf{u2} \\ \mathsf{v1} \\ \mathsf{v2} \\ \mathsf{v3} \end{array} \right) =
\left( \begin{array}{cccccc} 0 & 0 & 0 & 1 & 1 & 1 \\
  1 & 1 & 1 & 0 & 0 & 0 \\
  0 & 0 & 1 & 0 & 0 & 1 \\
  0 & 1 & 0 & 0 & 1 & 0 \\
  1 & 0 & 0 & 1 & 0 & 0
  \end{array}\right) \cdot
\left( \begin{array}{c} \mathsf{H1} \\ \mathsf{H2} \\ \mathsf{H3} \\ \mathsf{H4} \\ \mathsf{H5} \\ \mathsf{H6}  \end{array}\right),
\label{eq:example}
\end{equation}
or, more generally, in a matrix form:
\begin{equation}
y = Ax,
\label{eq:charge}
\end{equation}
where $y$ is a vector of charge measurements spanning the hit wires from all planes, $x$ is
a vector of expected charge in each possible hit cell to be solved, and $A$ is the biadjacency
matrix connecting wires and cells, which is determined solely by the wire geometry. We first consider the case where there is no uncertainty in the charge measurement and the bi-adjacency matrix $A$ is invertible. Upon solving Eq.~\ref{eq:charge}, the true hit cells will have the deduced charges equal to their true charges, while the fake hit cells will have the deduced charges equal to zero. In reality, the uncertainty in the charge measurement needs to be considered, and the $A$ matrix may not be invertible leading to ambiguous solutions. In the following, we describe detailed techniques in handling these challenges..

\paragraph*{Incorporating charge uncertainty: }
In practice, the charge measured on each wire has associated uncertainties from noise contributions
and imperfection of the signal processing procedures. Therefore, Eq.~\ref{eq:charge} 
is only approximate. The charge uncertainties can be taken into account by constructing a $\chi^2$ function:

\begin{eqnarray}\label{eq:chi2}
\chi^2
&=& (y-Ax)^T \cdot V^{-1} \cdot (y-Ax) \\ \nonumber
&\equiv& ||y'-A'x||_2^2,
\end{eqnarray}
where $V$ is the covariance matrix representing the uncertainty of the measured charge on each wire.
The vector $y$ and $x$ are then pre-normalized through $V^{-1} = Q^TQ$ (Cholesky decomposition),
$y' = Q \cdot y$, and $A' = Q \cdot A$. The notation $||\cdot||_p$ defines the $\ell_p$-norm of a
vector such that $||x||_p = (\sum_i|x_i|^p)^{1/p}$.

The best solution is found to be:
\begin{equation}\label{eq:solution}
x = \left( A^{T} \cdot V^{-1} \cdot A \right)^{-1} \cdot A^{T} \cdot V^{-1} \cdot y
\end{equation}
by minimizing the above $\chi^2$ function. Therefore, when the matrix
$M = A^{T} V^{-1} A$ is invertible, the best-fit charges
of hit cells can be derived directly using Eq.~\ref{eq:solution}.
The true hit cells will have the expected charges \emph{close to} the true charges,
while the fake hit cells will have the expected charges \emph{close to} zero.

\paragraph*{Incorporating sparsity: }
In most occasions, however, Eq.~\ref{eq:charge} is under-determined
and do not have a unique solution.  Similarly, the $\chi^2$ function constructed through
Eq.~\ref{eq:chi2} does not have a unique minimum location, representing a severe
loss of information due to the wire-readout system. In such cases, one would find that
the matrix $M = A^{T} V^{-1} A$ in Eq.~\ref{eq:solution} is non-invertible.

Additional constraints can be applied by considering the characteristics of typical physics events.
One such constraint is that most of the elements in the solution $x$ should be zero and so the solution should be \emph{sparse}.
In the example of Fig.~\ref{fig:ambiguity}, the sparse condition
implies that out of the 6 possible cells, many are fake. Then, Eq.~\ref{eq:charge} can be transformed into a constrained linear problem:
\begin{equation}
\text{minimize} \, ||x||_0, \quad \text{subject to:} \,\, y = Ax,
\label{eq:lp_l0}
\end{equation}
where $||x||_0$ is the $\ell_0$-norm of $x$, which represents the number of non-zero
elements in $x$. In other words, we seek the most sparse or simplest solution that explains the measurements. 
In the rare cases where the sparse condition fails to represent reality, the ability to resolve ambiguities from the wire-readout system would be intrinsically limited by the hardware.

Since the $\ell_0$-norm is non-convex, the optimization problem of Eq.~\ref{eq:lp_l0} is NP-hard~\cite{Natarajan}. Nonetheless, a practical procedure through combinatorial trials is described as follows:
\begin{enumerate}
	\item Calculate the number of zero eigenvalues $n_0$ of the matrix $M = A^{T} V^{-1} A$.
	\item Enumerate the $\binom{N}{n_0}$ possible reductions of Eq.~\ref{eq:charge} produced by removing $n_0$ elements from $x$ of original size $N$ and updating the $A$ and $V$ matrices accordingly.
	\item For each reduction, redo step 1. If $n_0$ is now zero,
          calculate a solution based on Eq.~\ref{eq:solution}, and
          record the $\chi^2$ value. Otherwise, go to the next
          reduction.
	\item Accept the solution with the minimal $\chi^2$ value.
\end{enumerate}

As an example, consider Fig.~\ref{fig:ambiguity}
and assume an identity covariance matrix.
The six eigenvalues of $6 \times 6$ matrix $M$ are 5, 3, 2, 2, 0 and 0. Thus there are two
zero eigenvalues. In order to reach a unique solution, two out of the six possible hit
cells need to be removed. 
The best solution is obtained by comparing $\chi^2$ values from the $\binom{6}{2} = 15$ combinations. One can see that this procedure, although sound in principle, becomes computationally intractable
when the hit multiplicity is more than a few tens. 


\paragraph*{Applying compressed sensing:}
Interestingly, the technique of \emph{compressed sensing}~\cite{cs}, originally proposed to
recover stable signal from incomplete and inaccurate measurements, solves the above computational
issues. Compressed sensing has wide applications in the field of electrical engineering~\cite{Donoho},
medicine and biology~\cite{cstomo},
applied statistics~\cite{l1statsmodel}, etc. Its application in high-energy and nuclear physics
experiments is relatively rare, but has great potential. The key concept is to approximate the
NP-hard $\ell_0$ problem of Eq.~\ref{eq:lp_l0} with a $\ell_1$ problem:
\begin{equation}
	\text{minimize} \, ||x||_1, \quad \text{subject to:} \,\, y = Ax,
	\label{eq:lp_l1}
\end{equation}
which retains the desired sparsity. The proof of this approximation can be found in Ref.~\cite{cs}.
Equivalently, the $\chi^2$ function defined in Eq.~\ref{eq:chi2} can be replaced by a $\ell_1$-regularized
$\chi^2$~\cite{lasso,bp} as follows:
\begin{equation}
\chi^2 = ||y'-A'x||_2^2 + \lambda ||x||_1,
\label{eq:chi2_l1}
\end{equation}
where $\lambda$ regulates the strength of the $\ell_1$-norm. Since the $\ell_1$-regularized $\chi^2$function is convex,
fast minimization can now be achieved through algorithms such as coordinate descent~\cite{coordesc}. In addition,
the \emph{non-negativity} physical constraint, defined as $x_i\geq0$, can also be incorporated during minimization.

With the application of the compressed sensing~\cite{cs}, the time complexity of the charge solving procedure is significantly
reduced from $\frac{n!}{m!(n-m)!}$ to O$(nk)$. Here, $n$, $m$, and $k$ are number of unknowns, number
of zero eigenvalues of the matrix $M$, and number of non-zero elements in $x$, respectively. Although it has
been proven that the solution of $\ell_1$-regularized $\chi^2$ is approaching the solution of the original
$\ell_0$ problem with a properly chosen regularization strength~\cite{cs}, some differences can exist given the
pre-chosen regularization strength. In practice, the overall performance of the charge solving procedure
can be evaluated in simulation by comparing the solution with the truth information. 

\begin{figure}
\centering
\includegraphics[width=0.8\textwidth]{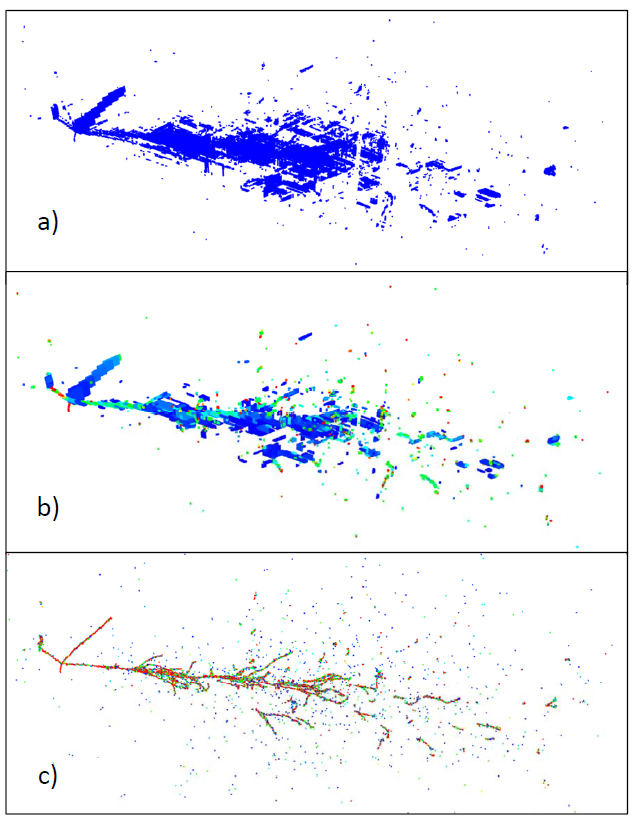}
\caption{\label{fig:quality} Comparison of 3D imaging quality with a) time and geometry information only; b) additional charge and sparsity information; and c) truth from Monte Carlo. The 3D web-based display can be found at \protect\url{http://www.phy.bnl.gov/wire-cell/bee/set/6/event/20/}. 
	See text for more discussions.  }
\end{figure}

\paragraph*{Example output:}
Compressed sensing substantially improves the computational speed and resolves the
scalability issue for high multiplicity events. Figure~\ref{fig:quality}
shows an example of a simulated 3-GeV $\nu_e$ charge current interaction, which has four tracks
including an electromagnetic shower exiting the primary vertex.
The 3D image is projected to a 2D view with the normal in the direction of the E-field. In this
projection, the ambiguity from isochronous condition is maximally displayed. The bottom panel
shows the ionizing energy deposition from Monte Carlo truth. The top panel shows the 3D reconstructed
image using only time and geometry information. All possible hit cells that are consistent with
the hit wires are shown. The middle panel shows the 3D reconstructed image after including
charge information and sparsity constraint. Comparing with the top panel, 35\% of fake cells are removed
in this example. The wrongly removed cells are below 1\%, which are mainly caused by finite charge resolution.
In addition, the imaging provides a measure of the inonization density, represented in the figure by color.
It is spread over local regions of remaining ambiguity such that total charge is conserved. More examples
can be found in Ref.~\cite{bee}.

\subsection{Other Considerations}
Given the basic principles of \wc{} 3D imaging described, we now discuss some details regarding implementation
for realistic detectors.

\paragraph*{Merging cells}
A single cell represents the finest spacial resolution defined by the wire
geometry.  In highly ambiguous time slices, considering the many individual
cells leads to a severely under-determined system. Consequently, the
$\ell_1$-problem of Eq.~\ref{eq:lp_l1} becomes susceptible to fluctuations in the
measurements, leading to mistakes in the solution. 
This problem can be mitigated by merging adjacent hit
cells into a single, larger cell while simultaneously merging their associating wires. Mathematically, the $y$ vector in Eq.~\ref{eq:charge} is replaced by $B \cdot y$, where
the matrix $B$ represents merging operation. The merging will reduce
the ambiguity at the expense of increasing the spacial resolution. Therefore, the extent of merging must be optimized using data to obtain the resolution necessary for physics events.

\paragraph*{Wrapped wires}
In experiments such as DUNE, some of the wires are wrapped around the anode planes, exposing them to signals from either side. Such a design is practical
for the ease of underground construction. However, it increases
the ambiguities already presented in wire-readout LArTPCs, as different cells may share the same wire measurements. The wrapped wires don't
need special treatment in the \wc{} procedures as long as the corresponding $x$
vector and $A$ matrix in Eq.~\ref{eq:charge} are constructed accordingly.
Meanwhile, the charge information used in \wc{} will reduce the additional ambiguities introduced by wrapped wires. 

\paragraph*{Dead channels}
In real experiments, 
dead wire channels could occur. In the 2D wire-versus-time views, they
manifest as gaps in continuous tracks, or missing hits in showers. Their effects
can be mitigated in \wc{} by treating the dead channels as always being hit. Consequently,
extra hit cells are generated including both true and fake ones. Ambiguities can
increase vastly near the dead channel region, causing ``ghost tracks''.
Typically, additional pattern recognition algorithms are necessary to identify and remove the ghosts, especially when the number of projective views are few. 

\paragraph*{Further constraints} The $\chi^2$ function defined in Eq.~\ref{eq:chi2_l1}
can be modified to include more constraints based on available knowledge, such as
connectivity, proximity, and so on. Such additional constraints typically
further reduce ambiguities, and add more robustness against fluctuations caused by inaccurate
measurements. Existing advanced algorithms such as \emph{group lasso}~\cite{l1statsmodel},
\emph{fused lasso}~\cite{fusedlasso}, and others, can be exploited.

The \wc{} techniques have been successfully applied to the MicroBooNE~\cite{Acciarri:2016smi}
experimental data. The quantitative evaluation of the detailed algorithm and
its performance is beyond the scope of current paper and will be reported in a future
publication~\cite{wc_uboone_imaging}.

\section{Conclusions}
The massive LArTPC is an advanced detector technology that may hold the key to many
future discoveries. In this paper, we present a novel 3D imaging method for
high-performance event reconstruction in large LArTPCs. Through simple and robust
mathematics, unique features of the LArTPC regarding the charge and sparsity information
are incorporated in addition to the
time and geometry to reduce ambiguities in image reconstruction.  The
3D image of ionization density from \wc{} significantly reduces the challenges in the later
pattern recognition and provides an excellent starting point for subsequent event reconstruction,
including detailed visual examination of the event. 
Further 3D fine tracking can be implemented through a combined fit of 3D clusters and their 2D
projections~\cite{Antonello:2012hu} and graph theory algorithms. Particle identification can be
developed through the $dE/dx$ information obtained in 3D. The 3D image also opens a new opportunity
for the generalized Convolutional Neural Network that has seen rapid development in recent LArTPC
applications based on 2D images~\cite{Acciarri:2016ryt}. In addition, techniques such as compressed
sensing used in \wc{} can be extended to other under-determined physics systems and are expected to
have a wide usage in the field of high-energy and nuclear physics.

\acknowledgments
This work is supported by the U.S. Department of Energy,
Office of Science, Office of High Energy Physics, and Early Career
Research Program under contract number DE-SC0012704.

\bibliographystyle{JHEP}
\bibliography{main}{}

\providecommand{\href}[2]{#2}\begingroup\raggedright\begin{thebibliography}{10}

\bibitem{pdg-nakamura}
K.~Nakamura and S.~Petcov, \emph{{in the Review of Particle Physics}},
  {\emph{Chin.Phys.} {\bfseries C40} (2016) 100001}.

\bibitem{Diwan:2016gmz}
M.~V. Diwan, V.~Galymov, X.~Qian and A.~Rubbia, \emph{{Long-Baseline Neutrino
  Experiments}},
  \href{http://dx.doi.org/10.1146/annurev-nucl-102014-021939}{\emph{Ann. Rev.
  Nucl. Part. Sci.} {\bfseries 66} (2016) 47--71},
  [\href{https://arxiv.org/abs/1608.06237}{{\ttfamily 1608.06237}}].

\bibitem{lartpc}
C.~Rubbia, ``The liquid-argon time projection chamber: A new concept for
  neutrino detector.'' CERN-EP/77-08 (1977).

\bibitem{lartpc_3}
H.~H. Chen, P.~E. Condon, B.~C. Barish and F.~J. Sciulli, ``{A Neutrino
  detector sensitive to rare processes. I. A Study of neutrino electron
  reactions}.'' FERMILAB-PROPOSAL-0496 (1976).

\bibitem{Willis:1974gi}
W.~Willis and V.~Radeka, \emph{{Liquid Argon Ionization Chambers as Total
  Absorption Detectors}},
  \href{http://dx.doi.org/10.1016/0029-554X(74)90039-1}{\emph{Nucl.Instrum.Meth.}
  {\bfseries 120} (1974) 221--236}.

\bibitem{Nygren:1976fe}
D.~R. Nygren, \emph{{The Time Projection Chamber: A New 4 pi Detector for
  Charged Particles}}, {\emph{eConf} {\bfseries C740805} (1974) 58}.

\bibitem{Anderson:2011ce}
{\scshape ArgoNeuT} collaboration, C.~Anderson et~al., \emph{{First
  Measurements of Inclusive Muon Neutrino Charged Current Differential Cross
  Sections on Argon}},
  \href{http://dx.doi.org/10.1103/PhysRevLett.108.161802}{\emph{Phys.Rev.Lett.}
  {\bfseries 108} (2012) 161802},
  [\href{https://arxiv.org/abs/1111.0103}{{\ttfamily 1111.0103}}].

\bibitem{Cavanna:2014iqa}
{\scshape LArIAT} collaboration, F.~Cavanna, M.~Kordosky, J.~Raaf and B.~Rebel,
  \emph{{LArIAT: Liquid Argon In A Testbeam}},
  \href{https://arxiv.org/abs/1406.5560}{{\ttfamily 1406.5560}}.

\bibitem{Berns:2013usa}
{\scshape CAPTAIN} collaboration, H.~Berns et~al., \emph{{The CAPTAIN Detector
  and Physics Program}},  in \emph{{Proceedings, 2013 Community Summer Study on
  the Future of U.S. Particle Physics: Snowmass on the Mississippi (2013)}}.
\newblock \href{https://arxiv.org/abs/1309.1740}{{\ttfamily 1309.1740}}.

\bibitem{Acciarri:2016smi}
{\scshape MicroBooNE} collaboration, R.~Acciarri et~al., \emph{{Design and
  Construction of the MicroBooNE Detector}},
  \href{http://dx.doi.org/10.1088/1748-0221/12/02/P02017}{\emph{JINST}
  {\bfseries 12} (2017) P02017},
  [\href{https://arxiv.org/abs/1612.05824}{{\ttfamily 1612.05824}}].

\bibitem{Amerio:2004ze}
{\scshape ICARUS} collaboration, S.~Amerio et~al., \emph{{Design, construction
  and tests of the ICARUS T600 detector}},
  \href{http://dx.doi.org/10.1016/j.nima.2004.02.044}{\emph{Nucl. Instrum.
  Meth.} {\bfseries A527} (2004) 329--410}.

\bibitem{Antonello:2015lea}
{\scshape MicroBooNE, LAr1-ND, ICARUS-WA104} collaboration, M.~Antonello
  et~al., \emph{{A Proposal for a Three Detector Short-Baseline Neutrino
  Oscillation Program in the Fermilab Booster Neutrino Beam}},
  \href{https://arxiv.org/abs/1503.01520}{{\ttfamily 1503.01520}}.

\bibitem{Acciarri:2016crz}
{\scshape DUNE} collaboration, R.~Acciarri et~al., \emph{{Long-Baseline
  Neutrino Facility (LBNF) and Deep Underground Neutrino Experiment (DUNE)}},
  \href{https://arxiv.org/abs/1601.05471}{{\ttfamily 1601.05471}}.

\bibitem{Qian:2015waa}
X.~Qian and P.~Vogel, \emph{{Neutrino Mass Hierarchy}},
  \href{http://dx.doi.org/10.1016/j.ppnp.2015.05.002}{\emph{Prog. Part. Nucl.
  Phys.} {\bfseries 83} (2015) 1},
  [\href{https://arxiv.org/abs/arXiv:1505.01891}{{\ttfamily
  arXiv:1505.01891}}].

\bibitem{Qian:2013ora}
X.~Qian, C.~Zhang, M.~Diwan and P.~Vogel, \emph{{Unitarity Tests of the
  Neutrino Mixing Matrix}},  \href{https://arxiv.org/abs/1308.5700}{{\ttfamily
  1308.5700}}.

\bibitem{Adams:2013qkq}
{\scshape LBNE Collaboration} collaboration, C.~Adams et~al., \emph{{The
  Long-Baseline Neutrino Experiment: Exploring Fundamental Symmetries of the
  Universe}},  \href{https://arxiv.org/abs/1307.7335}{{\ttfamily 1307.7335}}.

\bibitem{Radeka:2011zz}
V.~Radeka et~al., \emph{{Cold electronics for 'Giant' Liquid Argon Time
  Projection Chambers}},
  \href{http://dx.doi.org/10.1088/1742-6596/308/1/012021}{\emph{J. Phys. Conf.
  Ser.} {\bfseries 308} (2011) 012021}.

\bibitem{Acciarri:2017hat}
{\scshape MicroBooNE} collaboration, R.~Acciarri et~al., \emph{{The Pandora
  multi-algorithm approach to automated pattern recognition of cosmic-ray muon
  and neutrino events in the MicroBooNE detector}},
  \href{http://dx.doi.org/10.1140/epjc/s10052-017-5481-6}{\emph{Eur. Phys. J.}
  {\bfseries C78} (2018) 82},
  [\href{https://arxiv.org/abs/1708.03135}{{\ttfamily 1708.03135}}].

\bibitem{Acciarri:2017sde}
{\scshape MicroBooNE} collaboration, R.~Acciarri et~al., \emph{{Noise
  Characterization and Filtering in the MicroBooNE Liquid Argon TPC}},
  \href{http://dx.doi.org/10.1088/1748-0221/12/08/P08003}{\emph{JINST}
  {\bfseries 12} (2017) P08003},
  [\href{https://arxiv.org/abs/1705.07341}{{\ttfamily 1705.07341}}].

\bibitem{Baller:2017ugz}
B.~Baller, \emph{{Liquid argon TPC signal formation, signal processing and
  reconstruction techniques}},
  \href{http://dx.doi.org/10.1088/1748-0221/12/07/P07010}{\emph{JINST}
  {\bfseries 12} (2017) P07010},
  [\href{https://arxiv.org/abs/1703.04024}{{\ttfamily 1703.04024}}].

\bibitem{Adams:2018dra}
{\scshape MicroBooNE} collaboration, C.~Adams et~al., \emph{{Ionization
  Electron Signal Processing in Single Phase LArTPCs I. Algorithm Description
  and Quantitative Evaluation with MicroBooNE Simulation}},
  \href{https://arxiv.org/abs/1802.08709}{{\ttfamily 1802.08709}}.

\bibitem{Adams:2018gbi}
{\scshape MicroBooNE} collaboration, C.~Adams et~al., \emph{{Ionization
  Electron Signal Processing in Single Phase LArTPCs II. Data/Simulation
  Comparison and Performance in MicroBooNE}},
  \href{https://arxiv.org/abs/1804.02583}{{\ttfamily 1804.02583}}.

\bibitem{magnify_online}
 Online demonstration of signal processing: \url{http://lar.bnl.gov/magnify/}.

\bibitem{Li:2015rqa}
Y.~Li et~al., \emph{{Measurement of Longitudinal Electron Diffusion in Liquid
  Argon}}, \href{http://dx.doi.org/10.1016/j.nima.2016.01.094}{\emph{Nucl.
  Instrum. Meth.} {\bfseries A816} (2016) 160--170},
  [\href{https://arxiv.org/abs/1508.07059}{{\ttfamily 1508.07059}}].

\bibitem{Natarajan}
B.~K. Natarajan, \emph{Sparse approximate solutions to linear systems},
  \href{http://dx.doi.org/10.1137/S0097539792240406}{\emph{SIAM Journal on
  Computing} {\bfseries 24} (1995) 227--234}.

\bibitem{cs}
E.~J. Cand\`{e}s, J.~K. Romberg and T.~Tao, \emph{Stable signal recovery from
  incomplete and inaccurate measurements},
  \href{http://dx.doi.org/10.1002/cpa.20124}{\emph{Communications on Pure and
  Applied Mathematics} {\bfseries 59} (2006) 1207--1223},
  [\href{https://arxiv.org/abs/math/0503066}{{\ttfamily math/0503066}}].

\bibitem{Donoho}
D.~L. Donoho, \emph{Compressed sensing},
  \href{http://dx.doi.org/10.1109/TIT.2006.871582}{\emph{IEEE Transactions on
  Information Theory} {\bfseries 52} (2006) 1289--1306}.

\bibitem{cstomo}
H.~Yu and G.~Wang, \emph{Compressed sensing based interior tomography},
  {\emph{Physics in Medicine and Biology} {\bfseries 54} (2009) 2791}.

\bibitem{l1statsmodel}
M.~Yuan and Y.~Lin, \emph{Model selection and estimation in regression with
  grouped variables}, {\emph{Journal of the Royal Statistical Society. Series B
  (Statistical Methodology)} {\bfseries 68} (2006) 49--67}.

\bibitem{lasso}
R.~Tibshirani, \emph{Regression shrinkage and selection via the lasso},
  {\emph{Journal of the Royal Statistical Society. Series B (Methodological)}
  {\bfseries 58} (1996) 267--288}.

\bibitem{bp}
S.~S. Chen, D.~L. Donoho and M.~A. Saunders, \emph{Atomic decomposition by
  basis pursuit}, \href{http://dx.doi.org/10.1137/S1064827596304010}{\emph{SIAM
  J. Sci. Comput.} {\bfseries 20} (1998) 33--61}.

\bibitem{coordesc}
J.~Friedman, T.~Hastie and R.~Tibshirani, \emph{Regularization paths for
  generalized linear models via coordinate descent}, {\emph{Journal of
  statistical software} {\bfseries 33} (2010) 1--22}.

\bibitem{bee}
 Wire-Cell web-based 3D event display:
  \url{http://www.phy.bnl.gov/wire-cell/bee/}.

\bibitem{fusedlasso}
R.~Tibshirani, M.~Saunders, S.~Rosset, J.~Zhu and K.~Knight, \emph{Sparsity and
  smoothness via the fused lasso}, {\emph{Journal of the Royal Statistical
  Society. Series B (Statistical Methodology)} {\bfseries 67} (2005) 91--108}.

\bibitem{wc_uboone_imaging}
 "Application of Wire-Cell 3D Image Reconstruction in MicroBooNE", MicroBooNE
  collaboration, under preparation.

\bibitem{Antonello:2012hu}
M.~Antonello et~al., \emph{{Precise 3D track reconstruction algorithm for the
  ICARUS T600 liquid argon time projection chamber detector}},
  \href{http://dx.doi.org/10.1155/2013/260820}{\emph{Adv. High Energy Phys.}
  {\bfseries 2013} (2013) 260820},
  [\href{https://arxiv.org/abs/1210.5089}{{\ttfamily 1210.5089}}].

\bibitem{Acciarri:2016ryt}
{\scshape MicroBooNE} collaboration, R.~Acciarri et~al., \emph{{Convolutional
  Neural Networks Applied to Neutrino Events in a Liquid Argon Time Projection
  Chamber}},
  \href{http://dx.doi.org/10.1088/1748-0221/12/03/P03011}{\emph{JINST}
  {\bfseries 12} (2017) P03011},
  [\href{https://arxiv.org/abs/1611.05531}{{\ttfamily 1611.05531}}].

\end{thebibliography}\endgroup

\end{document}